# Correlation between structure and Rayleigh parameters in the lead-free piezoceramic (1-x)Ba(Ti$_{0.88}$ Sn$_{0.12}$)O$_3$-x(Ba$_{0.7}$Ca$_{0.3}$)TiO$_3$

Mulualem Abebe, Kumar Brajesh and Rajeev Ranjan*

Department of Materials Engineering, Indian Institute of Science Bangalore-560012, India

## Abstract

Composition dependent Rayleigh and structural analysis was carried out on the lead-free piezoceramics (1-x)(BaTi$_{0.88}$Sn$_{0.12}$)-x(Ba$_{0.7}$Ca$_{0.3}$) TiO$_3$ at room temperature. The system exhibits tetragonal (*P4mm*) structure for x > 0.21, rhombohedral (*R3m*) for x < 0.13 and orthorhombic (*Amm*2) for 0.13<x<0.21. Rayleigh analysis suggests that the irreversible contribution to the dielectric response is enhanced in the single phase orthorhombic compositions in the vicinity of the *R*3*m-Amm*2 and *Amm*2-*P*4*mm* phase boundaries, and not in compositions exhibiting phase coexistences (x = 0.12 and 0.22). We also found a correspondence between the irreversible Rayleigh parameter and the coercive field in this system.

*rajeev@materials.iisc.ernet.in

## I.    Introduction

In the past decade and half there is an increasing interest in research on Pb-free piezoelectrics [1-3]. An important breakthrough in this regard is the discovery of extraordinarily large piezoelectricity ($d_{33}$~620 pC/N) in a non-textured piezoceramics of (Ba,Ca)(Ti,Zr)O$_3$ [3]. Subsequent years have witnessed more systems exhibiting high piezo-response with Sn and Hf substitutions (in place of Zr) [4-9]. Alongside the development of new high-performance lead-free piezoelectrics, the scientific community is also engaged in understanding the origin of high piezoelectricity in such systems [10 - 19]. The polarization rotation theory lays emphasis on the anisotropic flattening of the free energy profile and availability of low energy pathways in the unit cell for the polarization vector to rotate [10 – 12] in a mono domain crystal. From the perspective of martensitic theory, the dominant contribution to dielectric and piezoelectric responses comes from the motion of the domain walls [13, 14]. Conventionally, the domain wall contribution to the dielectric and piezoelectric response is regarded as extrinsic, and that due to atomic displacements within the unit cell is regarded as intrinsic [15-19].

Rayleigh analysis one of the conventional approaches to analyze the contribution of domain wall motion with regard to the dielectric and piezoelectric response in piezoceramics [17]. This formalism is based on the movement of domain walls in a medium comprising of pinning centers (defects) or varying strengths [17, 18]. In the small signal regime, the motion of the domain walls is both reversible and irreversible in nature. The extent of the reversibility decreases as the coercive field is approached. The magnitude of the property (dielectric or piezoelectric) in the sub-coercive field regime is proportional to the amplitude of the field. Accordingly, the dielectric coefficient follows the relationship:

$$\varepsilon_r = \varepsilon_{rev} + \alpha_\varepsilon * E_o \qquad (1)$$

where $E_o$ is the amplitude of the cyclic sub-coercive electric field. $\varepsilon_{rev}$ and $\alpha E_o$ represent the reversible and the irreversible contributions, respectively to the measured dielectric response. The validity of this relationship has been tested for a wide variety of high performance piezoelectrics such as PZT [17-23], PbTiO$_3$-BiScO$_3$ [24], (Na,K)NbO$_3$-based system [25-28], and BaTiO$_3$ –based piezoceramics [29, 30].

Similar to (Ba,Ca)(Ti,Zr)O$_3$, the (Ba,Ca)(Ti, Sn)O$_3$ system also exhibits large piezoelectric response [4-6]. The first reported composition-temperature phase diagram of (Ba,Ca)(Ti, Sn) showed only tetragonal and rhombohedral ferroelectric phases [4]. Later studies however confirmed the presence of an intermediate orthorhombic (*Amm*2) phase region in between the tetragonal and rhombohedral phase fields [5]. A perusal of literature suggests that there is a lack of systematic correlation between the evolution of the Rayleigh parameters vis-à-vis the composition induced structural changes in modified BaTiO$_3$-based high performance piezoceramics. In the present study, we have carried out structural and Rayleigh analysis of the dielectric response as a function of composition, close to the polymorphic phase boundaries of a high performance lead-free piezoelectric system (1-x)(BaTi$_{0.88}$Sn$_{0.12}$) –x(Ba$_{0.7}$Ca$_{0.3}$)TiO3 at close composition intervals. We found that the system exhibits rhombohedral (R3m) for x<0.13, orthorhombic (Amm2) for 0.13< x < 0.21 and tetragonal (*P4mm*) for x > 0.21. In contrast to the lead-based piezoelectric systems such as PZT, the piezoelectric response in the BCTS was found to be maximum not for compositions exhibiting coexistence of ferroelectric phases, but for compositions exhibiting single orthorhombic phase in the vicinity of the *Amm*2-*P4mm* and *Amm*2-*R*3*m* phase boundaries. Our Rayleigh analysis of the P-E behaviour revealed a one-to-one correspondence between the enhancement in the piezoelectric response and the irreversible Rayleigh parameter for the dielectric permittivity. We also establish a correspondence between the irreversible Rayleigh parameter and coercive field in this system, and discuss the results in the light of other studies reported on Pb-free and Pb-based piezoelectrics.

## II. Experimental

Different compositions of (1-x) (BaTi$_{0.88}$Sn$_{0.12}$)-x(Ba$_{0.7}$Ca$_{0.3}$)TiO$_3$ with x = 0, 0.12, 0.13, 0.15, 0.21, 0.24, 0.27 and 0.29) were prepared by conventional solid-state reaction method. The starting materials were BaCO$_3$ (Alfa Aesar, with 99.8% purity), CaCO$_3$ (Alfa Aesar, with 99.95% purity), SnO$_2$ (Alfa Aesar, with 99.9% purity), and TiO$_2$ (Alfa Aesar, with 99.8% purity) were thoroughly mixed in zirconia jars with zirconia balls and acetone as the mixing medium using a planetary ball mill (Fritsch P5). The thoroughly mixed powder was calcined at 1100$^0$C for 4 hrs and milled again in acetone for 5hrs for better homogenization. The calcined powder was mixed with 2 wt. % polyvinyl alcohol (PVA) and pressed into disks of 15 mm diameter by using uniaxial dry pressing at 10 ton. Sintering of the pellets was carried out at 1300$^0$C for 4 hours and 1500 $^0$C for 6 hours under ambient condition. X-ray powder diffraction was done using a Rigaku (smart lab) with Johanson monochromator in the incident beam to remove the Cu-K$\alpha_2$ radiation. Measurement of the direct weak-field longitudinal piezoelectric coefficient (d$_{33}$) was carried using piezotest PM 300 by poling the pellets at room temperature for 1h at a field of ~2.2 kV/mm. Polarization electric field (P-E) hysteresis loop were measured with a Precision premier II loop tracer. Structure refinement was carried out using FULLPROF software [31].

### III.  Results

#### A. *Piezoelectric and structural studies*

Fig. 1 shows the P-E hysteresis loops of (1-x) BST – x BCT for different compositions. The well-defined loops with saturated polarization confirm the ferroelectric nature of these compositions. The weak-field longitudinal piezoelectric coefficient $d_{33}$ exhibits a sharp increase with increasing x in the composition range 0<x<0.13. The rate of increase of $d_{33}$ with composition is significantly reduced in the composition range 0.13<x<0.21, reaching a maximum of 475 pC/N at x= 0.21. For x > 0.21, $d_{33}$ drops slightly. Fig. 2 shows the compositional evolution of representative pseudo cubic x-ray Bragg profiles. Consistent with the anomalous changes in the properties at x = 0.13 and x = 0.21, a notable change in the shape of the Bragg profiles can be seen at the two boundary compositions. The splitting of the $\{400\}_{pc}$ profile, with nearly equal intensity for x = 0.13 is consistent with orthorhombic (Amm2) distortion of the perovskite cell [8]. The Bragg profiles remain nearly unchanged in the composition range 0.13<x<0.21. This is consistent with the nearly constant $d_{33}$ in the same composition range, Fig 1b. For x ≥ 0.22, the nature of the $\{400\}_{pc}$ splitting changes abruptly. Not only the separation between the peaks increases, but also the intensity ratio is close to 1:2. This suggests onset of a tetragonal (P4mm) distortion. This transformation is accompanied by a slight decrease in $d_{33}$ at x > 0.21, Fig1b. We carried out Rietveld fitting of the diffraction patterns with different structural models. The diffraction patterns of the intermediate compositions 0.13≤x≤0.21 could be fitted satisfactorily with a single phase *Amm*2 structure. The compositions in the immediate vicinity of this region namely x =0.12 and x=0.22 were best fitted with Amm2 + R3m and P4mm + Amm2 phase coexistence models, respectively. For sake of consistency, the atomic coordinates of the structures in the two phase refinement were borrowed from the compositions exhibiting single phase. The refined structural parameters of compositions exhibiting single phase tetragonal, orthorhombic and rhombohedral structures are given in Table I.

#### B. *Rayleigh analysis*

Rayleigh analysis was carried out on the cyclic electric field (E) polarization (P) data obtained in the sub-coercive field regime. Fig.4a shows typical P-E curves for representative compositions in the sub coercive field regime (0.2 kV/cm to 0.8 kV/cm). The dielectric coefficient was calculate obtained using the expression

$$\varepsilon^* = P_{p-p}/2E_o \qquad (2)$$

Where $P_{p-p}$ is peak-to-peak polarization measured for applied electric field amplitude $E_o$. The linearity of $\varepsilon_r - E_o$ confirms the validity of the Rayleigh approach in the chosen field region, fig. 4b. The reversible $\varepsilon_{rev}$ and the irreversible α Rayleigh parameters were obtained by fitting the linear $\varepsilon_r - E_o$ plots with eq. 1. The composition dependence of the two coefficients is plotted in Fig. 5a. The reversible parameter $\varepsilon_r$ increases sharply from ~1200 to ~1500 when the composition increased from x=0.11 to 0.13. This variation less in the composition range 0.13≤x≤0.21. $\varepsilon_r$ decreases again for x > 0.21. The irreversible component α on the other hand exhibits a two distinct peaks at x=0.13 and 0.21. We can define a term *f$_{irrev}$* as the fraction of the irreversible contribution to the measure dielectric response

$$f_{\text{irrev}} = \frac{\varepsilon_{\text{irreversible}}}{\varepsilon_{\text{reversible}}+\varepsilon_{\text{irreversible}}} = \frac{\alpha_\varepsilon E_o}{\varepsilon_{\text{rev}}+\alpha_\varepsilon E_o} \qquad (3)$$

From the composition dependence of *f$_{irrev}$* shown in Fig. 5b, it is evident that the irreversible contribution is 4% − 7% for the rhombohedral (*R3m*) compositions, slightly higher ( 11%) for the tetragonal (*P4mm*) composition, and signficantly large (~ 24%) for the orthorhombic (*Amm*2) compositions.

### IV.    Discussion

Our results show that although the reversible contribution to the dielectric coeffieicnt varies slowly in the orthorhombic phase region 0.13≤x≤0.21, the irreversible parameter α exhibits clear maxima at the orthorhombic compositions x=0.13 and 0.21 in the immediate vicinity of the orthorhombic-rhombohedral and orthorhombic-tetragonal phase boundaries, respecctively. These results corroborate the enhanced extrinsic contribution at the orthorhombic-tetragonal phase boundary reported earlier Acosta et al [30] for the (Ba,Ca)(Ti, Zr)$O_3$ system. The fraction of the irreversible contribution is maximum (*f$_{irrev}$* ~ 24 %, Fig. 5b) at x=0.21. The f$_{irrev}$ in the BCTZ system is ~ 18 % (See Table 2) for a tetragonal composition in the vincinity of the *P4mm-Amm*2 phase boundary. This correlates well with the enhanced motion of tetragonal ferroelectric-ferroelastic domain walls in the BCTZ system as measured by *in-situ* electric field dependent high energy XRD [32]. The corresponding tetragonal composition in the vicinity of the *P4mm-Amm*2 transition in our system is x=0.22, Fig.3. However, this composition exhibits reduced value of the irreversible Rayleigh parameters and piezoelectric coefficient, Fig.1b. Another contrasting feature is that the rhombohedral compositions of BCTZ exhibit higher $f_{irrev}$ than the orthorhombic (which is depicted as MPB in Ref. [29]) which is not the situation in our case. The inconsistencies between our results and those reported for BCTZ suggest the need for a detailed comparative study of the correlations between structure and Rayleigh parameters in the two $BaTiO_3$-based systems. It is worth comparing our results with the modified $K_{0.5}Na_{0.5}NbO_3$ (KNN) based piezoelectric systems which exhibit the same sequence of structural transitions as the $BaTiO_3$-based piezoelectrics. Huan et al [26] reported that the irreversible Rayleigh parameter ($\alpha$) exhibits maximum in the orthorhombic composition in the immediate vicinity of the orthorhombic-tetragonal phase boundary of $(Na_{0.52}K_{0.4425}Li_{0.0375})(Nb_{0.92-x}Ta_xSb_{0.08})O_3$. The irreversible Rayleigh parameter was also found to exhibit maximum for an orthorhombic composition in the vicinity of P4mm-Amm2 boundary in another KNN based piezoelectric $(Na_{0.5}K_{0.5})_{1-x}Li_xNbO_3$ [27]. Peng et al carried out a temperaure dependent structural and Rayleigh analysis of $0.95(K_{0.5}Na_{0.5})NbO_3$-$0.05LiTaO_3$ across the *P4mm-Amm*2 transition and reported the irreversible contribution to be most enhanced at the temperature exhibiting single phase orthorhombic structure in the vicinity of the phase coexistence region [25]. These results are similar to ours. From the reversible ($\varepsilon_{rev}$) and irreversible ($\alpha$) Rayleigh parameters reported by Huan et al [26], we estimated that irreversible contribution to the total dielectric response to be $f_{irrev}$ ~ 40 % in $(Na_{0.52}K_{0.4425}Li_{0.0375})(Nb_{0.92-x}Ta_xSb_{0.08})O_3$. This value is considerably larger than the irreversible contribution exhibited by our system ($f_{irev}$ ~ 24 %), and also by the BCTZ ($f_{irrec}$ ~ 18 %, see Table II). Peng *et al*. have also reported that reversible and the irreversible Rayleigh parameters exhibit maximum at different temperatures. The reversible Rayleigh parameter peaks at ~ 105 $^0$C, whereas the irreversible Rayleigh parameter exhibits maximum at ~ 70 $^0$C [25]. In contrast, our composition dependent study shows that both the reversible and the reversible Rayleigh parameters follows the same trend with composition, Fig. 5a. The different trend of the reversible and irreversible Rayleigh parameter may indicate that the (reversible) lattice contribution to the total dielectric response starts to dominate the system over the contributions resulting from the domain wall motion. This lattice contribution can become signficant when the system approaches a critical point. In $BaTiO_3$ based piezoelectrics, the convergence of the rhombohedral-orthorhombic, orthorhombic-tetragonal and tetragonal-cubic phase boundaries occurs ~ 50 – 60 $^0$C [33-35]. The relative permittivity shows anomlaous increase when the system approaches this covergence region. In this scenario the reversible

Rayleigh parameter ($\varepsilon_{rev}$) is likely to exhibit maximum at a different composition/temperature than the irreversible Rayleigh parameter.

Since the convergence region also includes the cubic paraelectric, the polarization of the system is likely to decay when the convergence region is approached. As a consequence, although the dielectric response increases abruptly, the piezoelectric response is expected to decrease in the close vicinity of convergence region [36]. In view of this, a large value of reversible Rayleigh ($\varepsilon_{rev}$) parameter need not correate with a good piezoelectric response. Rather, a one-to-one correspondence should be sought between composition/temperature dependene of the piezoelectric response and the composition/temperature dependence of the irreversible Rayleigh parameter ($\alpha$). In this context, we may anticipate a correlation between the coercive field and the irreversible Rayleigh parameter since coercive field is a measure of the ease of domain switching in ferroelectric materials, which the irreversible Rayleigh parameter represents. We found that although $\alpha$ exhibits a highly non-monotonic composition dependence (two peaks at x =0.13 and 0.21, Fig 5a and 5b), the coercive field and the extrinsic Rayleigh parameter ($\alpha$) shows a monotonic dependence, Fig. 5c. The irreversible Rayleigh parameter $\alpha$ in the present case therefore corresponds to the ease of domain switching.

While there is a general tendency to associate the irreversible Rayleigh parameter ($\alpha$) with the irreversibility of the domain wall motion in piezoceramics, studies in the last few years have shown that ferroelectric systems in the vicinity of interferroelectric phase boundaries exhibit field induced phase transformation [37-43]. Hinterstein et al [39], have argued that field induced interferroelectric transformation have dominant contribution to the electrostrain observed in the MPB compositions of PZT. This view found support from Khatua *et al.* in another analogous MPB system (1-x)PbTiO$_3$-(x)BiScO$_3$ (PT-BS) [40]. High energy synchrotron x-ray diffraction study of this system *in-situ* with electric field has revealed that the switching propensity of the rhombohedral domain is considerably reduced in the MPB composition as compared to the rhombohedral composition just outside the MPB. The high piezoelectric response of the MPB composition PT-BS was attributed to the field induced rhombohedral-tetragonal transformation and the associated interphase boundary motion. Khatua et al [41] have shown that the domain wall motion and phase transformations are highly correlated phenomenon in the MPB composition of (1-x) PbTiO$_3$-xBiScO$_3$. In such a scenario, the Rayleigh coefficients would represent a complex interrelated motion of domain wall motion and interphase boundaries. The fact that maximum piezoelectric response in our system occurs in the single phase orthorhombic (*Amm*2) compositions in the immediate vicinity of the *P*4*mm*-*Amm*2 and *R*3m-*Amm*2 boundaries appears to contrast with the Pb-based piezoelectrics wherein enhanced piezoelectric response is often reported to be exhibited by compositions showing coexistence of ferroelectric phases. In view of our results, it would be worth examining if a similar correspondence exist between coercive-field, irreversible Rayleigh parameter and piezoelectric response in the lead-based systems or not. Such as comparative study will help in determining if the correlation reported by us is a universal feature across different MPB systems or it is system specific.

## V. Conclusions

In summary, we investigated the correlation between structural states and the Rayleigh parameters for the dielectric coefficient in the lead-free piezoelectric ceramic (1-x)(BaTi$_{0.88}$Sn$_{0.12}$)–x (Ba$_{0.7}$Ca$_{0.3}$)TiO$_3$. We found that the compositions exhibiting single phase orthorhombic structure in the vicinity of the *P*4*mm*-*Amm*2 and *R*3*m*-*Amm*2 phase boundaries phase show significantly enhanced irreversible Rayleigh parameter α, and not those exhibiting coexistence of ferroelectric phases. We also found that irrespective of the non-monotonic composition dependence of α, there exists a monotonic correspondence between the coercive field and the irreversible Rayleigh parameter in this system.

**Acknowledgements**  Kumar Brajesh gratefully acknowledges the Science and Engineering Research Board (SERB) of the Department of Science and Technology, Govt. of India for the award of National Post Doctoral Fellowship. R. Ranjan is grateful to the SERB for financial assistance (Grant No. EMR/2016/001457) and for financial assistance to the IISc-STC cell.

**References**

[1]  Y. Saito, H. Takao, T. Tani, T. Nonoyama, K. Takatori,  T. Homma, T. Nagaya, M. Nakamura *Nature* **432**  84 (2004)

[2] J. Rödel, W. Jo, K. T. P. Seifert, E.-M. Anton, T. Granzow, and D. Damjanovic, J. Am. Ceram. Soc. **92**, 1153 (2009).

[3] W. Liu and X. Ren, Phys. Rev. Lett. **103**, 257602 (2009)

[4] Y. Yao, C. Zhou, D. Lv, D. Wang, H. Wu, Y. Yang, and X. Ren, Europhys. Lett. 98, 27008 (2012).

[5] D. Xue, Y. Zhou, H. Bao, J. Gao, C. Zhou and X. Ren, Appl. Phys. Lett. 99, 122901 (2011)

[6] C. Zhou, W. Liu, D. Xue, X. Ren, H. Bao, J. Gao, and L. Zhang, Appl. Phys. Lett. **100**, 222910 (2012)

[7] A. K. Kalyani, K. Brajesh, A. Senyshyn and R. Ranjan, Appl. Phys. Lett. **104**, 252906 (2014)

[8] A. K. Kalyani, H. Krishnan, A. Sen, A. Senyshyn, and R. Ranjan, Phys. Rev. B **91**, 024101 (2015)

[9] L.-F. Zhu, B.-P. Zhang, X.-K. Zhao, L. Zhao, F.-Z. Yao, X. Han, P.-F. Zhaou, and J.-F. Li, Appl. Phys. Lett. **103,** 072905 (2013)

[10] H. Fu, R. E. Cohen, Nature **403** 281 (2000)

[11] D. Vanderbilt and M.H. Cohen, Phys. Rev. B **63**, 094108 (2001)

[12] D. Damjanovic, J. Am. Ceram. Soc. 88, 2663 (2005)

[13] Y. M. Jin, Y. U. Wang, A. G. Khachaturyan, J. F. Li, D. Viehland, J. Appl. Phys. **94,** 3629 (2003)

[14] Y.M. Jin, Y.U. Wang, A.G. Khachaturyan, J.F. Li, and D. Viehland, Phys. Rev. Lett. **91**, 197601 (2003).

[15] S. Li, W. Cao, and L. E. Cross, J. Appl. Phys. 69, 7219 (1991)

[16] J. Y. Li, R. C. Rogan, E. Üstündag, and K. Bhattacharya, Nature Mater. **4**, 776 (2005)

[17] D. Damjanovic, Rep. Prog. Phys. 61, 1267 (1998)

[18] D. Damjanovic, M. Demartin, J. Phys. D: Appl. Phys. 29, 2057 (1996)

[19] D. A. Hall, J. Mat. Sc. 36, 4575 (2001)

[20] S. Trolier-McKinstry, N. B. Gharb, and D. Damjanovic, Appl. Phys. Lett.88, 202901 (2006)

[21] J. E. Garcia, R. Perez, D. A. Ochoa, A. Albareda, M. H. Lente, J. A. Eiras, J. Appl. Phys. 103, 054108 (2008)

[22] D. –J. Kim, J. –P. Maria, A. I. Kingon, S. K. Streiffer, J. Appl. Phys. 93, 5568 (2003)

[23] A. Pramanick, D. Damjanovic, J. C. nino, J. L. Jones, J. Am. Ceram. Soc. 92, 2291 (2009)

[24] R. E. Eitel, T. R. Shrout, C. A. Randall, J. Appl. Phys. 99, 124110 (2006)

[25] B. Peng, Z. Yue, L. Li, J. Appl. Phys. 109, 054107 (2011)

[26] Y. Huan, X. Wang, L. Li, and J. Koruza, Appl. Phys. Lett. 107, 202903 (2015)

[27] K. Kobayashi, K. Hatano, Y. Mizuno, and C. A. Randall, Appl. Physics Express 5, 031501 (2012)

[28] D. A. Ochoa, G. Esteves, J. L. Jones, F. R. Marocs, J. F. Fernandez, and J. E. Garcia, APL, 108, 142901 (2016).

[29] J. Gao, X Hu, Le Zhang, F. Li, L. Zhang, Y. Wang, Y Hao, L. Zhang, and X. Ren, Appl. Phys. Lett. 104, 252909 (2014)

[30] M. Acosta, N. Novak, G. A. Rossetti Jr., and J. Roedel, Appl. Phys. Lett. 107, 142906 (2015)

[31] Rodrigues-J. Carvajal. FullPROF 2000 A Rietveld Refinements and Pattern Matching Analysis Program. France: Laboratories Leon Brillouin (CEA-CNRS)

[32] G. Tutuncu, B. Li, K. Bowman, and J. L. Jones, J. Appl. Phys. 115, 144104 (2014)

[33] D. S. Keeble, F. Benabdallah, P. A. Thomas, M. Maglione, and J. Kreisel, Appl. Phys. Lett. **102**, 092903 (2013)

[34] A. K. Kalyani, K. Brajesh, A. Senyshyn and R. Ranjan, Appl. Phys. Lett. 104, 252906 (2014).

[35] Z. Yu, C. Ang, R. Guo, and A. S. Bhalla, J. Appl. Phys. 92, 1489 (2002).

[36] M. Acosta, N. Khakpash, T. Someya, N. Novak, W. Jo, H. Nagata, G. A. Rossetti, Jr., J. Roedel, Phys. Rev. B. **91**, 104108 (2015)

[37] Lalitha K V, A. N. Fitch and R. Ranjan, Phys. Rev. B **87**, 064106 (2013)

[38] T. Iamsasri, G. Tutuncu, C. Uthaisar, S. wongsaenmai, S. Pojprapai, J. L. Jones, J. Appl. Phys. 117, 024101 (2015)

[39] M. Hinterstein, M. Hoelzel, J. Rouquette, J. Haines, J. Glaum, H. Kungl, and M. Hoffman, Acta Mater. **94**, 319 (2015)

[40] D. K. Khatua, Lalitha K. V., C. M. Fancher, J. L. Jones, and R. Ranjan, Phys. Rev. B 93, 104103 (2016).

[41] D. K. Khatua, Lalitha K. V., C. M. Fancher, J. L. Jones, and R. Ranjan, J. Appl. Phys. 120, 154104 (2016).

[42] A. K. Kalyani, H. Krishnan, A. Sen, A. Senyshyn, and R. Ranjan, Phys. Rev. B **91**, 024101 (2015)

[43] K. Brajesh, M. Abebe and R. Ranjan, Phys. Rev. B **94**, 104108 (2016).

Table I. Refined structural parameters of orthorhombic (x=0.21), tetragonal (x=0.27) and rhombohedral (x=0.05) phases of (1-x) BST- x BCT.

| a)  x=0.21 | | Space group: *Amm*2 | | |
|---|---|---|---|---|
| Atoms | x | y | z | B(Å$^2$) |
| Ba/Ca | 0.000 | 0.000 | 0.000 | 0.10 (3) |
| Ti/Sn | 0.500 | 0.000 | 0.512 (6) | 0.26 (4) |
| O1 | 0.000 | 0.000 | 0.532(3) | 0.11(3) |
| O2 | 0.500 | 0.192 (3) | 0.282 (3) | 0.10 (6) |
| a= 4.004 1(4) Å, b= 5.6699 (8) Å, c= 5.6752(7) Å | | | | |

| b)  x=0.27 | | Space group: *P4mm* | | |
|---|---|---|---|---|
| Atoms | x | y | z | B(Å$^2$) |
| Ba/Ca | 0.000 | 0.000 | 0.000 | 0.37 (3) |
| Ti/Sn | 0.500 | 0.000 | 0.516 (2) | 0.22 (3) |
| O1 | 0.500 | 0.500 | -0.002(2) | 0.23(7) |
| O2 | 0.500 | 0.000 | 0.464 (3) | 0.73 (8) |
| a= 4.0009(5) Å,  c= 4.0142(6) Å | | | | |

| c)  x=0.05 | | Space group: *R3m* | | |
|---|---|---|---|---|
| Atoms | x | y | z | B(Å$^2$) |
| Ba/Ca | 0.000 | 0.000 | 0.000 | 0.10 (4) |
| Ti/Sn | 0.000 | 0.000 | 0.494 (1) | 0.18 (5) |
| O | 0.349(8) | 0.203 (3) | 0.674 (7) | 0.16 (9) |
| a= 5.6820(5) Å, c= 6.9611(11) Å | | | | |

**Table II** Rayleigh coefficient for (1-x)BST-xBCT. The error estimated from the least squares analysis is below each coefficient. The extrinsic contribution is estimated by eq. (5) using $E_o = 0.8 kV/cm$.

| Compositions(x) | 0.11 | 0.12 | 0.13 | 0.15 | 0.21 | 0.22 | 0.24 | KNN-based | PZT | BS-PT | BCTZ |
|---|---|---|---|---|---|---|---|---|---|---|---|
| Crystal structure | R3m | R3m | Amm2 | Amm2 | Amm2 | P4mm | P4mm | Amm2 | MPB | MPB | PPB |
| $\alpha_\varepsilon$ (x10$^{-3}$ m/V) | 2.12 | 2.18 | 4.88 | 2.99 | 5.91 | 2.22 | 2.36 | | - | - | |
| $\varepsilon_{rev}$ | 1199 | 1332 | 1539 | 1532 | 1518 | 1191 | 1131 | | - | - | |
| Extrinsic contribution ($f_{irrev}$) (%) | 12.3 | 11.6 | 20 | 13.5 | 24 | 15.6 | 16.7 | 38.4[a], ~40[b] | 22.4[c] | 25.4[d] | ~18[e] |

[a] Ref [27], [b] Ref [26], [c] Ref [18], [d] Ref [24], [e] value estimated from data provided in Ref [29].

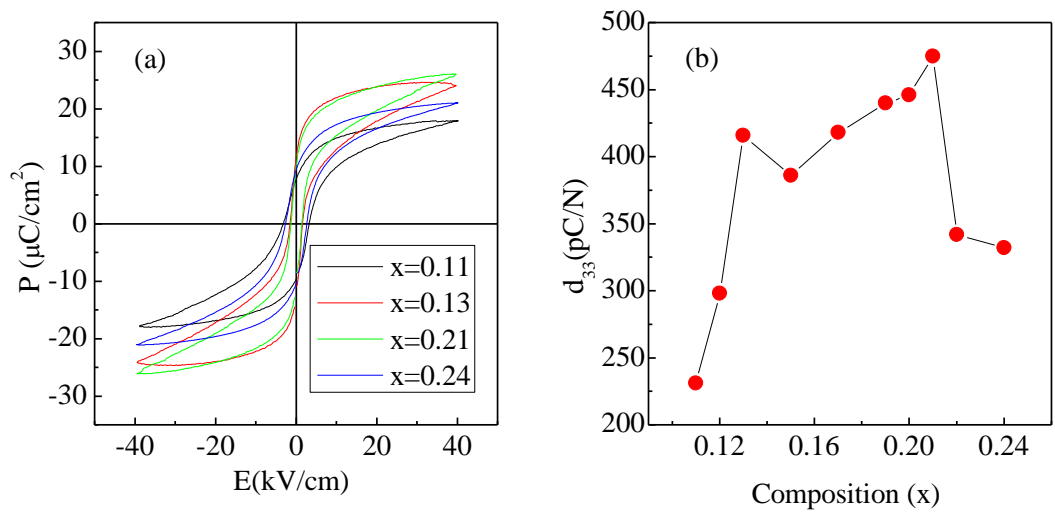

Fig. 1a shows the P-E hysteresis loops of (1-x) BST – x BCT for different composition (x). (b) shows composition variation of the weak-field direct longitudinal piezoelectric coefficient ($d_{33}$).

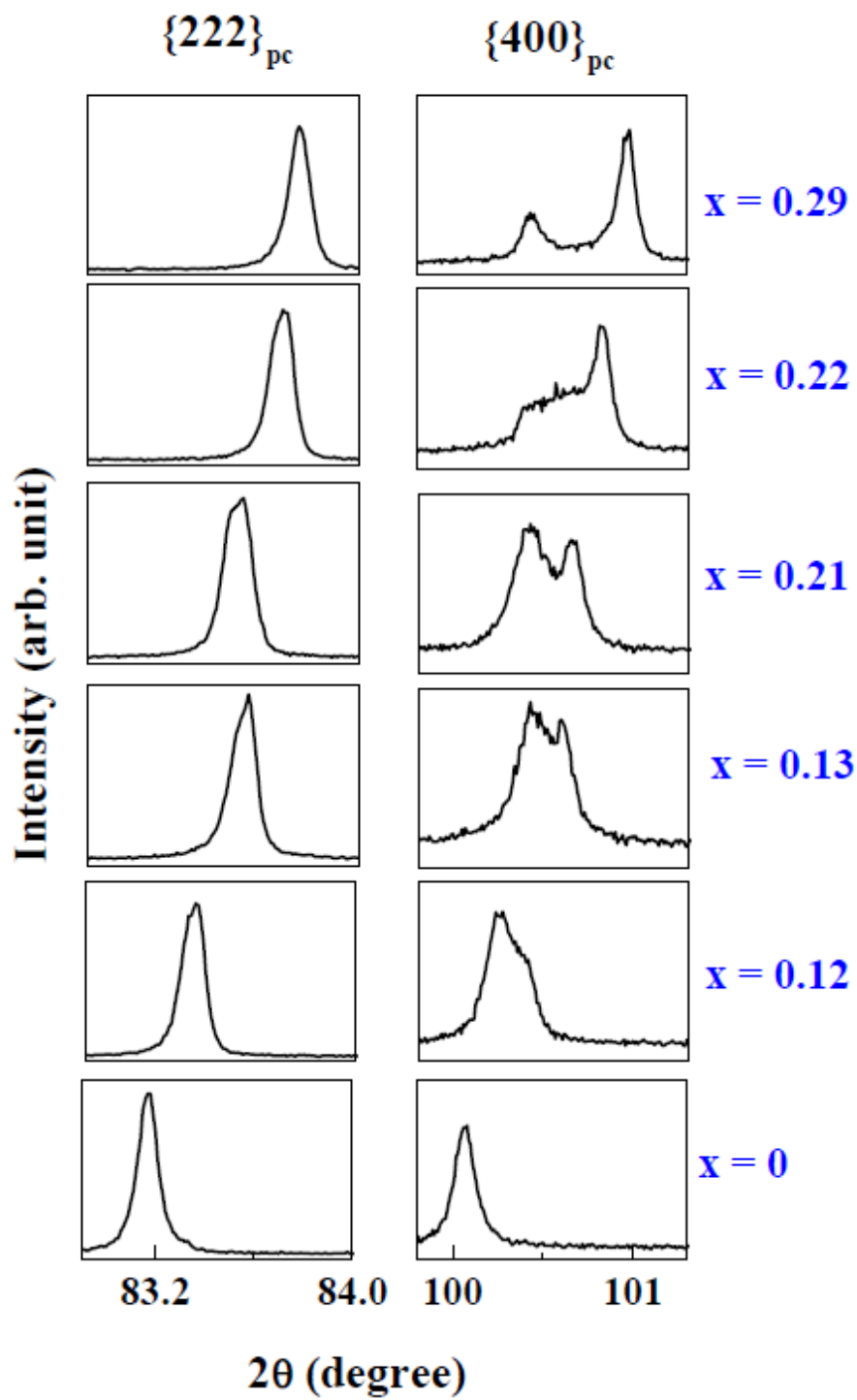

Fig. 2 X-ray Bragg profiles of the (1-x) BST-x BCT for x = 0, 0.12, 0.13, 0.21, 0.22, and 0.29.

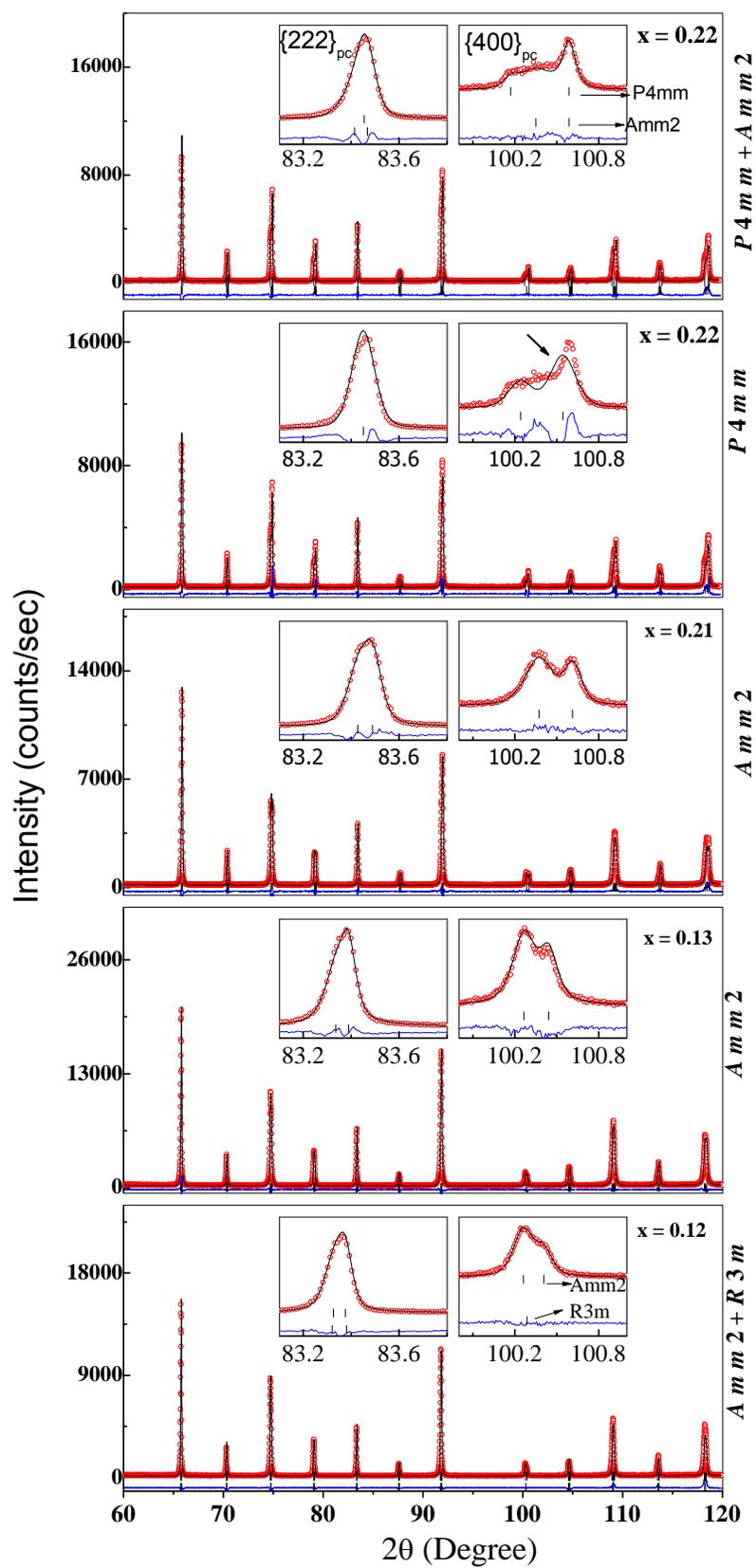

Fig. 3 Rietveld fitted XRD patterns of (a) x =0.12, 0.13, 0.21 and 0.22. For sake of comparison we have shown that the pattern of x =0.22 does not fit well with single phase P4mm structural model (second figure from top). It required the inclusion of the orthorhombic (Amm2) structure to account for the features not accounted by the pure P4mm model (shown by arrow).

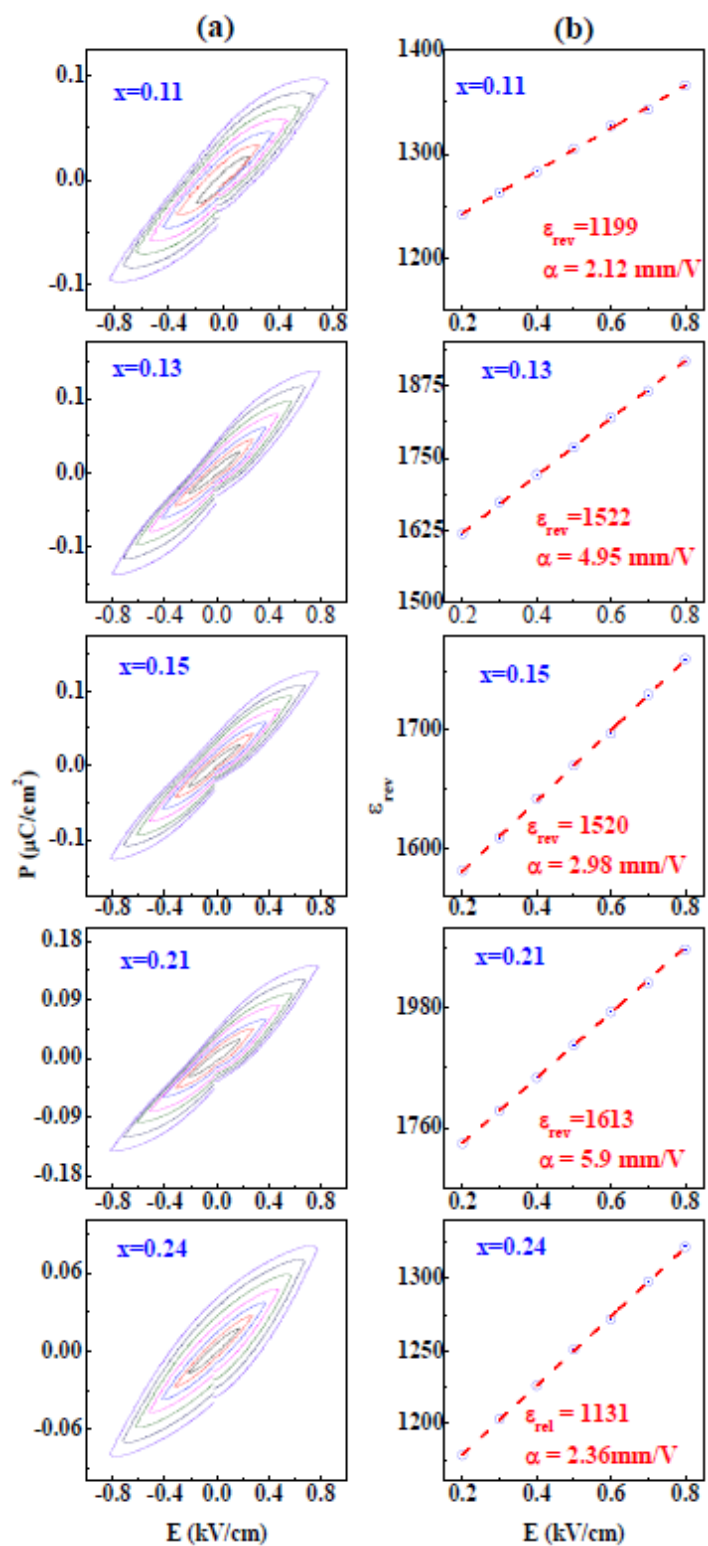

Fig. 4 (a) shows the polarization-electric field loop and (b) the corresponding fitted results of Rayleigh parameters at room temperature with a serious of electric field strength from 0.2kV/cm to 0.8kV/cm.

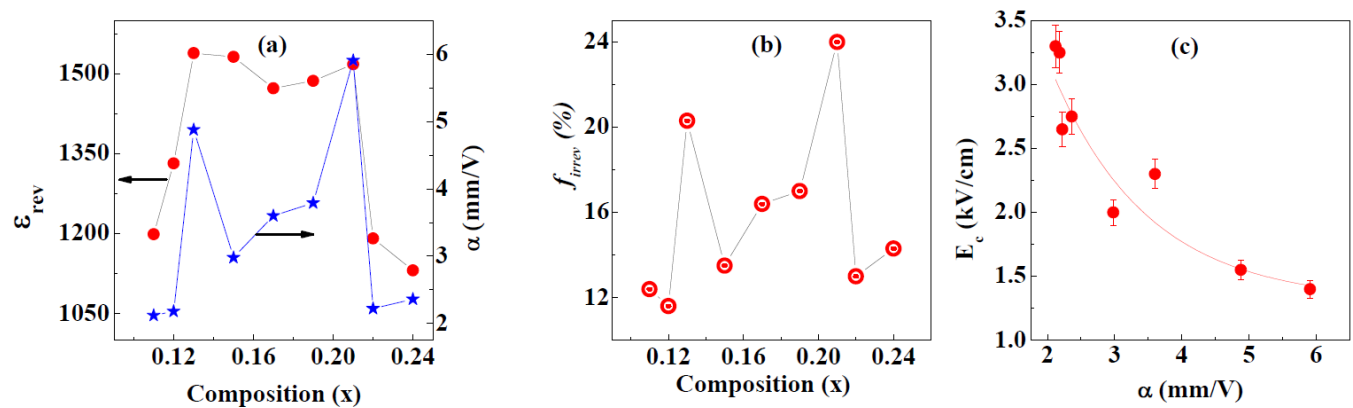

Fig. 5 (a) Reversible and irreversible Rayleigh parameters as a function of composition. (b) shows the composition variation of the fraction of the irreversible contribution to the measure dielectric response. (c) Shows the variation of coercive field as a function of the irreversible Rayleigh coefficient.